\documentclass[sigconf, screen]{acmart}

\usepackage{booktabs}
\usepackage{multirow} 
\usepackage{pifont}
\usepackage{graphicx} %
\usepackage{makecell}
\usepackage{adjustbox}
\usepackage{forest}
\usepackage{subcaption}
\usepackage{wrapfig}
\usepackage{tablefootnote}
\usepackage{threeparttable}
\usepackage[most]{tcolorbox}

\usepackage{colortbl}
\usepackage{color}
\usepackage{xcolor}
\usepackage{makecell}
\usepackage{enumitem}
\usepackage{graphicx}
\usepackage[frozencache]{minted}
\usepackage{xcolor}

\usepackage{microtype}
\definecolor{lightTeal}{HTML}{add2ca}
\definecolor{softPurple}{HTML}{8c89ba}
\definecolor{paleGreen}{HTML}{e6f0db}
\definecolor{mutedOrange}{HTML}{dfa660}
\definecolor{limeGreen}{HTML}{97c04b}
\definecolor{lightBlue}{HTML}{dbeef3}
\definecolor{softRed}{HTML}{d77470}
\definecolor{dustyBlue}{HTML}{93b2d9}
\definecolor{tabletitle}{HTML}{E5DBF8} 

\definecolor{checkgreen}{HTML}{009900}
\definecolor{crossred}{HTML}{CC0000}

\AtBeginDocument{%
  }

\newif\ifhjjmark
\hjjmarkfalse   

\newcommand{\hjj}[1]{\ifhjjmark{\color{softRed}#1}\else#1\fi}

\newcommand{\onlineappendix}{online appendix}

\AtBeginDocument{%
  }

\setcopyright{cc}
\setcctype{by}  
\copyrightyear{2026}
\acmYear{2026}
\acmDOI{10.1145/3832783.3837457}

\acmConference[ASE '26]{Proceedings of the 41st IEEE/ACM International Conference on Automated Software Engineering}{October 12--16, 2026}{Munich, Germany}
\acmISBN{979-8-4007-2882-2/2026/10}
\acmBooktitle{Proceedings of the 41st IEEE/ACM International Conference on Automated Software Engineering (ASE '26), October 12--16, 2026, Munich, Germany}

\begin{document}

\title{RealisticTritonBench: A Benchmark for Triton-Kernel Generation in Real-World AI Frameworks}

\author{Jinjun Huang}
\email{huanghuanghuang@zju.edu.cn}
\affiliation{%
  \institution{College of Computer Science and Technology and the State Key Laboratory of Blockchain and Data Security, Zhejiang University}
  \city{Hangzhou}
  \country{China}
}

\author{Zhongzhen Wen}
\email{wenzhongzhen@smail.nju.edu.cn}
\affiliation{
  \institution{State Key Lab for Novel Software Technology, Nanjing University}
  \city{Nanjing}
  \country{China}
}

\author{Tongtong Xu}
\email{xutongtong9@huawei.com}
\affiliation{
  \institution{Software Engineering Application Technology Laboratory, Huawei}
  \city{Hangzhou}
  \country{China}
}

\author{Meng Yan}
\email{mengy@cqu.edu.cn}
\affiliation{
  \institution{The School of Big Data and Software Engineering, Chongqing University}
  \city{Chongqing}
  \country{China}
}

\author{Xin Xia}
\email{xin.xia@acm.org}
\affiliation{
  \institution{College of Computer Science and Technology and the State Key Laboratory of Blockchain and Data Security, Zhejiang University}
  \city{Hangzhou}
  \country{China}
}

\author{Zhongxin Liu}
\authornote{Corresponding author.}
\email{liu\_zx@zju.edu.cn}
\affiliation{
  \institution{College of Computer Science and Technology and the State Key Laboratory of Blockchain and Data Security, Zhejiang University}
  \city{Hangzhou}
  \country{China}
}

\renewcommand{\shortauthors}{J. Huang, Z. Wen, T. Xu, M. Yan, X. Xia, and Z. Liu}

\begin{abstract}
In modern AI frameworks, the GPU kernel is a key determinant of overall system performance. By combining usability, portability, and near-handwritten CUDA performance, Triton has been widely adopted for implementing GPU kernels. 
Recent advances have shown the potential of using large language models (LLMs) to automatically generate Triton kernels, helping reduce the manual effort required by expert kernel developers. 
To evaluate the quality of LLM-generated Triton kernels, several benchmarks have been proposed. However, existing benchmarks primarily target isolated kernel generation tasks and suffer from three key limitations: (1) they restrict the task to only PyTorch-to-Triton translation, thus failing to reflect the diversity and complexity of real-world Triton tasks; (2) they evaluate only the performance of individual kernels, limiting the evaluation of real-world performance of generated kernels in AI frameworks, where end-to-end performance is the core criterion for real deployment; (3) they rely on manually written evaluation scripts for single kernel, which may introduce vulnerabilities and allow models to exploit evaluation flaws to bypass correctness checks and obtain inflated scores.
\hjj{To address these limitations, we introduce RealisticTritonBench, the first benchmark that derives Triton kernel generation tasks from real-world pull requests in popular AI frameworks, enabling evaluation under realistic, production-like settings.}
RealisticTritonBench systematically extracts real-world PRs with Triton kernel modified from popular open-source AI frameworks and transforms them into kernel generation tasks with concrete engineering contexts. Each task takes a natural language requirement as input and requires to generate a corresponding Triton kernel implementation. RealisticTritonBench also provides a complete and reproducible evaluation environment for each task.  
In contrast to prior benchmarks that focus on isolated kernel performance, RealisticTritonBench integrates generated kernels into their original frameworks and evaluates them under end-to-end test, enabling a more faithful assessment. 
We conduct a systematic evaluation of the state-of-the-art LLMs on RealisticTritonBench, revealing that they still struggle to handle the challenges in real-world Triton generation tasks.
\end{abstract}



\maketitle

\section{INTRODUCTION}

Currently, modern AI models are typically deployed on training and inference frameworks, such as Pytorch~\cite{paszke2019pytorchimperativestylehighperformance}, and Deepspeed~\cite{deepspeed2020}. 
Within these frameworks, GPU kernels constitute a critical component that determines system latency and throughput, motivating engineers to invest significant effort in developing high-performance kernels, as inefficient implementations can increase latency, reduce throughput, and raise the overall cost of large-scale model deployment~\cite{hsu2024liger}.
However, the implementation of correct and highly optimized GPU kernels using low-level native languages such as CUDA is inherently challenging and time-consuming~\cite{Triton2019}. 
To address these challenges, Triton~\cite{Triton2019} has emerged as a widely adopted high-level GPU programming language. It has been prevalently integrated into modern AI frameworks and optimization pipelines, serving as a preferred choice for implementing high-performance GPU kernels~\cite{kwon2023efficient,hu2025lightllm,zheng2024sglangefficientexecutionstructured}. 
As a Python-based domain-specific language (DSL) for GPU kernel development, Triton significantly simplifies the implementation of complex kernels while maintaining performance competitive with native CUDA. 
As shown in Figure~\ref{fig:example_triton}, a Triton kernel typically resembles a Python function decorated with @triton.jit, where developers explicitly define block-level parallelism, memory access patterns, and tensor computation through Triton primitives. 
Owing to its ease of use, flexibility, and strong performance, Triton has gained substantial traction in modern AI systems and has become a key building block for optimized critical kernels in contemporary AI frameworks. 
However, such DSLs have not fully eliminated the complexities of performance tuning. 
Developers are still required to reason about low-level execution details, including memory access patterns, parallelization strategies, and hardware-specific optimizations. Hence, current research in automated Triton generation has attracted increasing attention. 

\begin{figure}[t]
\centering
\begin{minipage}{1.0\linewidth}
\begin{minted}[
fontsize=\small,
bgcolor=gray!10,
linenos=false,
breaklines=true,
breakanywhere=true,
breaksymbolleft=,
breaksymbolright=,
breakindent=2em
]{python}
# An Example for Vector Addition
@triton.jit
def add_kernel(x_ptr, y_ptr, output_ptr,  n_elements, BLOCK_SIZE: tl.constexpr):
    pid = tl.program_id(axis=0)  
    block_start = pid * BLOCK_SIZE
    offsets = block_start + tl.arange(0, BLOCK_SIZE)
    mask = offsets < n_elements
    x = tl.load(x_ptr + offsets, mask=mask)
    y = tl.load(y_ptr + offsets, mask=mask)
    output = x + y
    tl.store(output_ptr + offsets, output, mask=mask)
\end{minted}
\end{minipage}
\caption{A Triton kernel for Vector Addition}
\label{fig:example_triton}
\end{figure}

More recently, a line of work has explored leveraging large language models (LLMs) to automatically synthesize Triton kernels. Several benchmarks have been proposed to evaluate models’ capabilities of generating Triton kernels\cite{ouyangkernelbench, wen2025multikernelbenchmultiplatformbenchmarkkernel, li2025tritonbenchbenchmarkinglargelanguage, xing2026flashinfer}. 
These benchmarks provide an initial testbed for assessing the functional correctness and performance of generated kernels, helping LLM-assisted kernel generation rapidly progress. Representative approaches, including AutoTriton\cite{li2025autotritonautomatictritonprogramming}, QiMeng-Kernel \cite{zhu2026qimeng}, and KernelFalcon \cite{kernelfalcon2025}, investigate LLM-based GPU kernel generation and reveal the potential of LLMs for automating Triton kernel generation. 
However, we find these key limitations in existing benchmarks: (1) Limited task diversity. Existing benchmarks predominantly focus on translating PyTorch reference implementations into Triton kernels \cite{li2025tritonbenchbenchmarkinglargelanguage, ouyangkernelbench, xing2026flashinfer}. In contrast, the real-world Triton generation task encompasses a broader range of activities, including performance optimization, bug fixing, and feature extension, as these tasks typically involve modifying existing Triton kernel implementations to meet new functional or performance requirements. These practical development scenarios remain insufficiently explored in existing studies. 
(2) Lack of framework-level evaluation. Existing benchmarks often evaluate kernels in isolation, only considering kernel-level metrics such as functional correctness and speedup compared to PyTorch implementations\cite{li2025tritonbenchbenchmarkinglargelanguage, ouyangkernelbench}. 
Such kernel-level evaluation overlooks the real impact of deploying generated kernels within AI frameworks. 
In practice, Triton kernels must interact with complex runtime systems, memory management strategies, distributed execution pipelines, and framework-specific abstractions. 
Without end-to-end evaluation in real-world AI frameworks, it is difficult to assess their genuine effectiveness and robustness with kernel-level evaluation. 
(3) Vulnerability in evaluation pipeline. Most benchmarks rely on manually written evaluation scripts\cite{li2025tritonbenchbenchmarkinglargelanguage, ouyangkernelbench}, which may introduce vulnerabilities and allow models to exploit evaluation flaws to obtain misleading results. Some researchers have observed that models may exploit flaws in evaluation scripts to bypass correctness checks, thereby obtaining inflated scores\cite{reddit_sakana_cheating_2025,liu2026dr}.

To address these limitations, we present RealisticTritonBench to enable evaluation under realistic environments. 
RealisticTritonBench has several characteristics: 1) Diverse tasks: we construct RealisticTritonBench from real-world Triton-related pull requests, covering a wide range of development activities, including performance optimization, modification, and new kernel addition, rather than limiting to PyTorch-to-Triton translation tasks. 
2) Multi-dimensional evaluation: To better approximate real-world deployment scenarios, we provide kernel-level unit tests, model accuracy tests, and end-to-end speed tests for tasks whenever possible. By incorporating end-to-end accuracy and performance evaluation, our benchmark directly measures the real impact of generated kernels within complete frameworks. 
3) Mitigate Hacking in Evaluation. Since end-to-end metrics directly reflect the model's actual performance, improvements at this level correspond to actual performance gains. As a result, it becomes difficult to exploit evaluation flaws without genuine improvements in system performance, thereby reducing the risk of misleading results.

We evaluate a diverse set of state-of-the-art LLMs on RealisticTritonBench, such as Qwen3.5 and GPT-5.4. To systematically measure task completion, we design these metrics for comprehensive evaluation: full test pass rate (FTP), unit test pass rate (UTP), numerical robustness for the model (NR), and end-to-end acceleration speedup. Together, these metrics form a realistic, system-level evaluation framework for generated kernels. To ensure comprehensive evaluation, our end-to-end testing pipeline is applied across a diverse set of models, covering different architectures and capabilities, summarized in our online appendix\cite{huang2026realistictritonbench}. Based on these metrics, we systematically study the performance, robustness, and failure reasons of LLMs on RealisticTritonBench. Our experiments show that models achieve only 18.71\% average task success rate, maintain model accuracy in 47.65\% of cases, and present limited end-to-end speedup (approximately 1× on average), which indicates that generating correct Triton kernels in real-world tasks remains challenging, and the generated implementations often lead to degraded model accuracy or increased end-to-end latency.

In summary, this paper makes the following contributions:

\begin{itemize}[leftmargin=*,topsep=1em]
\item We curate the RealisticTritonBench dataset, providing a foundation for evaluating the performance of AI-generated kernels in realistic environments.
\item We propose a comprehensive evaluation pipeline for Triton kernel generation, which spans unit testing to end-to-end system-level testing, closely simulating real-world kernel validation procedures.
\item We conduct an extensive evaluation of state-of-the-art language models and LLM-based agents on RealisticTritonBench, revealing that even the SOTA LLMs still struggle with real-world Triton kernel generation tasks, and the generated kernels frequently lead to degraded model accuracy and increased end-to-end latency.
\end{itemize}

\section{BACKGROUND}

\subsection{LLM Training and Inference Frameworks}

Recently LLMs have been widely applied to a variety of tasks including code generation\cite{cao2026qwen3}, mathematical reasoning\cite{shao2025deepseekmath}, video generation\cite{wiedemer2025video}, and multimodal understanding\cite{li2025llava}. As model scale and capability continue to grow, efficiently training and deploying these large-scale models has become increasingly significant. To address this challenge, researchers and engineers have developed various training and inference frameworks. For example, in the training task, frameworks such as PyTorch\cite{paszke2019pytorchimperativestylehighperformance}, Tensorflow\cite{abadi2016tensorflow} DeepSpeed\cite{deepspeed2020}, and Megatron-LM\cite{shoeybi2019megatron} provide essential capabilities including automatic differentiation, gradient optimization, fault tolerance, and large-scale distributed training. Although general-purpose deep learning frameworks such as PyTorch and TensorFlow are widely used for LLM inference, they are primarily designed to support a broad range of hardware platforms and model architectures. As a result, they do not specifically optimize the core decoding process of LLM inference, which may lead to suboptimal performance and inefficient resource utilization.

To improve inference efficiency, a number of specialized LLM inference engines have recently been developed\cite{kwon2023efficient, zheng2024sglangefficientexecutionstructured, zhang2025efficient,ggerganov2023llamacpp}. These systems introduce a variety of optimizations tailored for LLM inference and serving, including techniques such as dynamic batching\cite{ali2020batch} and inference-specific operators\cite{ye2025flashinfer}. By incorporating these optimizations, modern inference engines provide features such as efficient batching, optimized attention computation, and distributed inference, enabling LLMs to be deployed and executed more efficiently on GPU hardware.

\subsection{Triton for High-Performance GPU Kernel Development}

As deep learning models continue to grow in scale, GPU computational efficiency has become a key factor affecting both training and inference performance. In GPU programming, a GPU kernel refers to a function that runs on the GPU and is executed in parallel by a large number of threads, typically performing data-parallel computations on tensors or matrices. These kernels constitute the fundamental building blocks of many operations in deep learning systems.
Although general deep learning frameworks such as PyTorch provide a rich set of built-in operators, many performance-critical computations in real-world systems still rely on customized GPU kernels to achieve higher efficiency. 
For example, in scenarios like attention computation, matrix operations, and sparse computation, kernels optimized for specific data layouts and memory access patterns can significantly reduce latency.

Traditionally, high-performance GPU kernels are implemented using CUDA, Nvidia’s parallel computing platform\cite{wen2025multikernelbenchmultiplatformbenchmarkkernel}. However, writing efficient CUDA kernels is complex and requires deep expertise in GPU architecture, including thread scheduling and memory management. To reduce this burden while retaining high performance, Triton\cite{Triton2019} has emerged as a domain-specific language for GPU programming. It provides a Python-based abstraction that enables developers to write high-level kernels, while the compiler handles low-level optimizations such as parallelization and memory access. By combining high-level programmability with performance close to hand-written CUDA kernels, Triton significantly reduces the development complexity of high-performance GPU operators. As a result, Triton has been increasingly adopted in modern deep learning systems to implement GPU kernels. For example, systems such as vLLM\cite{kwon2023efficient} and libraries such as FlashAttention\cite{dao2022flashattention} use Triton to implement high-performance GPU kernels in their core components. However, implementing efficient Triton kernels still requires substantial effort for developers, which motivates efforts to explore the use of large language models for automatic kernel generation.

\section{RealisticTritonBench}

\begin{figure}[t]
\centering
\includegraphics[width=1.0\linewidth]{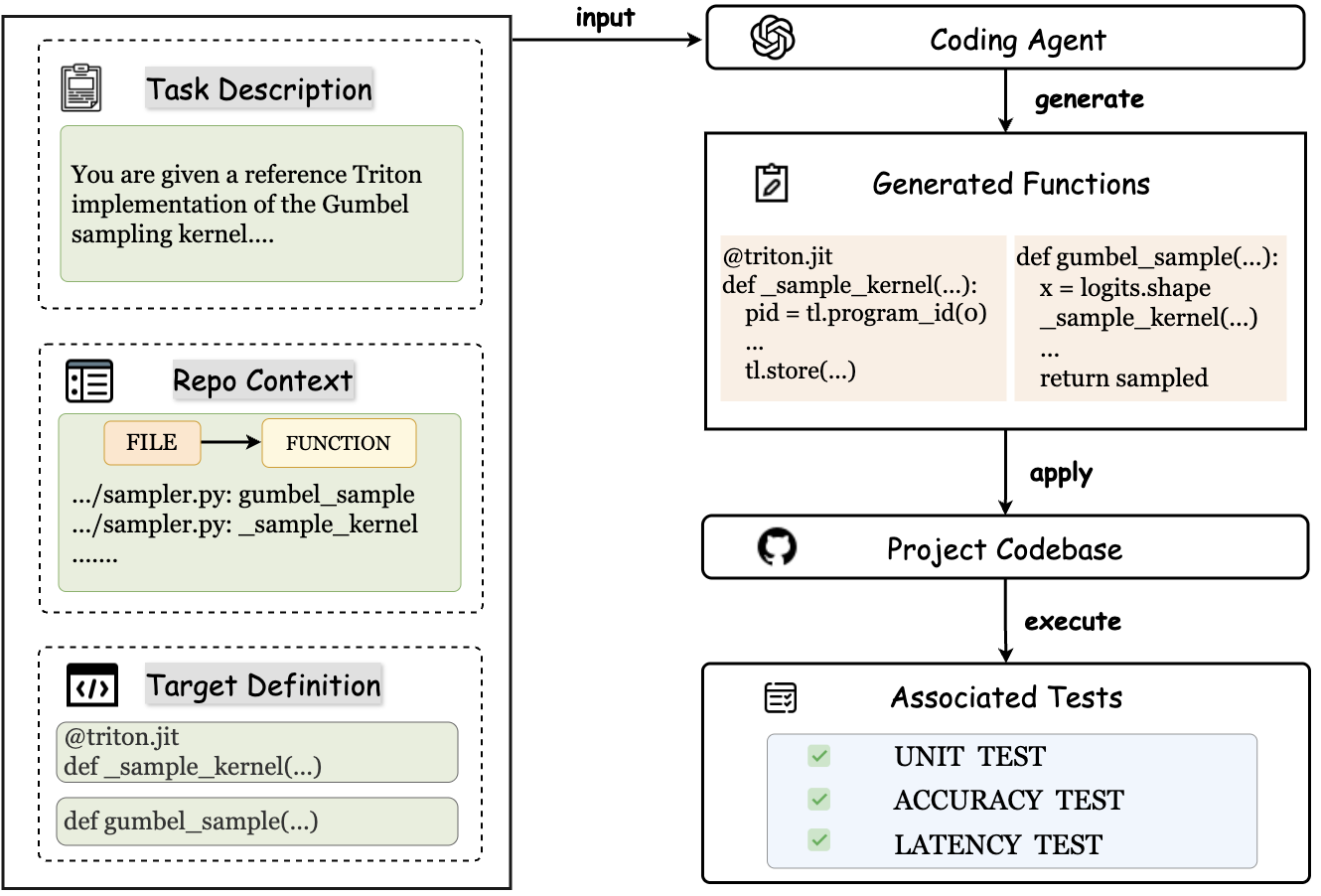}
\caption{Task formulation and evaluation pipeline of RealisticTritonBench}
\label{fig:overview}
\end{figure}

RealisticTritonBench aims to evaluate the ability of large language models (LLMs) to generate Triton kernels in realistic development scenarios. The overview of the benchmark is illustrated in Figure~\ref{fig:overview}. The benchmark consists of 31 real Triton kernel tasks collected from popular open-source AI frameworks. Each task corresponds to a merged pull request (PR) related to Triton kernel.

For each task, RealisticTritonBench provides three components as input: a task description, relevant context, and the definition of the target function. These components are combined to form the prompt that is provided to the LLM. LLM is required to implement the target function strictly following the specified requirements.

After generation, the produced kernels replace the original implementation in the prepared testing environment. We then execute a standardized evaluation pipeline that includes kernel-level unit tests, model-level accuracy tests, and end-to-end test. Specifically, the unit tests are first executed to verify the functional correctness of the generated kernels. The modified kernels are then integrated into the framework to evaluate model accuracy and end-to-end latency. The evaluation pipeline ensures that generated kernels satisfy the expected behavior of the task, while also capturing their impact on model accuracy and system-level performance under realistic deployment settings.

In this section, we present the task formulation and the construction pipeline of RealisticTritonBench in detail.

\subsection{Task Formulation}

\subsubsection{\textbf{Input Definition and Target Output}} To evaluate the ability of models to generate Triton kernels in real-world development environment, we construct the task inputs to closely resemble the information available to developers during kernel implementation. Specifically, each task consists of three components: task description, context, and target function specification.

The task description provides a detailed specification of the kernel functionality, describing the intended behavior and implementation constraints of the operator. The context contains relevant information extracted from the original repository, presented in the form of file-function and file–class references. These references indicate functions and classes that may be useful for implementing the target kernel, enabling the model to access the necessary information required for development.
    
The definition of the target function  defines the Triton kernel interface that the model is expected to implement. Given the task description and the relevant context, the model must generate a Triton kernel that conforms exactly to the provided  definition.

\subsubsection{\textbf{Evaluation Suite}}

Compared with previous benchmarks that focus only on evaluation at the single-operator level, we summarize practical development experience from real-world PRs and design three types of tests to improve the robustness of the benchmark: Unit Test, Model Accuracy Test, and Latency Test. 

Unit Test is used to verify the functional correctness of kernels. Model Accuracy Test evaluates whether replacing the operator implementation leads to noticeable degradation in model performance on common tasks. Latency Test measures the end-to-end system latency after the substitution.

\subsection{Benchmark Construction}

\begin{figure*}[t]
\centering
\includegraphics[width=0.95\linewidth]{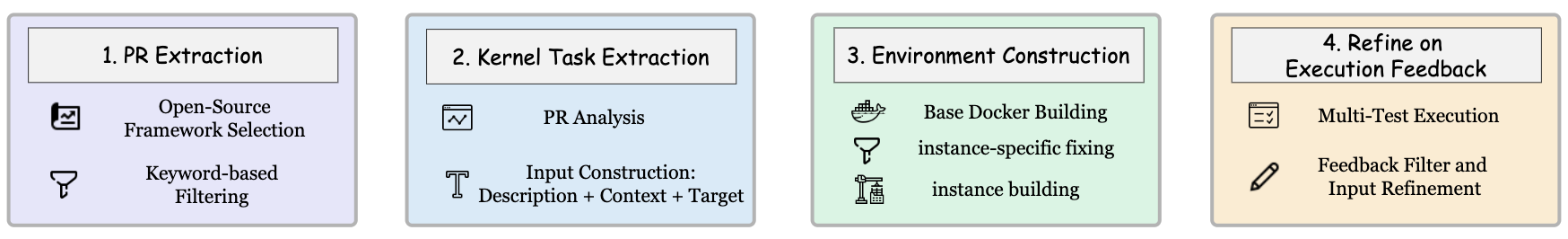}
\caption{The pipeline of constructing our benchmark}
\label{fig:pipeline}
\end{figure*}

We focus on constructing high-quality, real-world Triton kernel generation tasks and build RealisticTritonBench through a four-phase pipeline. As illustrated in Figure~\ref{fig:pipeline}, we first collect relevant pull requests from  selected open-source AI framework repositories (Phase 1). We then filter and analyze these PRs to extract concrete Triton kernel generation tasks (Phase 2). Next, we construct executable evaluation environments for each instance(Phase 3). Finally, we validate and refine the instances based on execution results to ensure the correctness and stability of the benchmark (Phase 4).

\subsubsection{PR Collection}

The goal of this phase is to collect pull requests (PRs) related to Triton kernel modifications from open-source repositories, in order to construct a candidate task pool grounded in real-world engineering practice. We first select several popular and actively maintained AI framework repositories as data sources, including PyTorch, vLLM, and SGLang. These repositories extensively use Triton to implement high-performance GPU kernels for both training and inference pipelines, making them representative of real-world Triton development scenarios. We therefore collect historical pull requests (PRs) from these repositories as candidate instances for further analysis.

To efficiently identify Triton-related modifications among a large number of PRs, we design an automated filtering strategy as an initial filter. Specifically, we first perform a coarse-grained filtering step based on keywords appearing in PR titles, descriptions, and commit messages, such as \textit{triton}, \textit{kernel}, \textit{operation}, and \textit{optimization}. This step helps quickly locate PRs that are potentially related to Triton. Next, we further analyze the code diffs in each PR and retain only those that actually modify or introduce Triton kernel implementations, such as files containing \texttt{@triton.jit} kernel definitions. By combining keyword-based filtering with code-level change analysis, our approach reduces manual checking effort while maintaining high relevance to Triton kernel modifications.

Through this process, we obtain a set of PRs closely related to Triton kernel modifications as candidate instances. These PRs cover a variety of real development scenarios, including kernel optimizations, bug fixes, and the implementation of new features, providing a diverse foundation for constructing the benchmark.

\subsubsection{Kernel Task Extraction}

Based on the PRs related to Triton kernel modifications collected in the previous stage, we conduct a detailed analysis of each PR to identify the objectives of the kernel changes, and subsequently transform them into well-defined Triton kernel generation tasks.

\noindent \textbf{Initial Filtering}. After the initial keyword-based filtering in the previous step, we obtained approximately 2,000 PRs that require manual analysis. We then manually examine each PR and further filter them according to the following evaluation criteria.
\noindent \begin{enumerate}[left=0pt,topsep=0em]
\item \textbf{Triton Kernel Relevance.}  
The PR must contain modifications related to Triton kernels.
    
\item \textbf{Test Availability.}  
The PR must include corresponding tests (e.g. at least unit tests) that can be used to verify correctness.
    
\item \textbf{Clear Kernel Objective.}  
The modification introduces a clear functional or performance-related objective that can be formulated as a standalone kernel task.
\end{enumerate}
    
\noindent \textbf{Task Extraction.} To extract tasks from PRs, we first conduct a manual analysis of each PR by carefully examining the PR description and the corresponding code changes. Through this process, we aim to understand the background, objectives, and motivations behind the Triton kernel modifications. For example, some PRs introduce new Triton kernels to replace existing PyTorch implementations in order to improve performance, while others may extend the functionality of existing kernels. This analysis allows us to clearly identify the target functions involved and the expected functional changes introduced by the modification.

Based on the content of the PR, we then employ an LLM to automatically generate the corresponding kernel task description. 
The generated description is typically summarizing the target function to be implemented or modified and the expected behavior. 
In this way, real-world code modifications are abstracted into executable Triton kernel generation tasks. 
Then, the generated descriptions are manually reviewed and revised when necessary to ensure that they accurately reflect the original intention of the PR and provide a clear and complete specification of the task.
\hjj{Specifically, each task description is checked for intent faithfulness to the PR goal, requirement completeness regarding behavior, constraints, and success criteria, and the absence of implementation leakage from the gold patch. 
Two Triton-experienced developers inspect the PR description, diff, target function, and tests, and score each criterion on a 0/1/2 scale.
If any criterion averages below 1.5, they revise the task description to consensus.
}
In addition to the task description, we also manually analyze and extract the contextual information required for implementing the task, including related functions and classes , the definition of the target function to be implemented, and the corresponding tests. We categorize the tests into three types: Unit Test, Model Accuracy Test, and Latency Test. The Unit Test relies on existing pytest-based testing commands provided by the repository. For Model Accuracy Test and Latency Test, we analyze the models associated with each PR and construct testing commands based on the repository’s existing evaluation scripts. These tests are used to verify model accuracy after kernel replacement and to measure the end-to-end system latency respectively.

\subsubsection{Environment Construction}

Following SWE-bench\cite{jimenez2024swebench}, we provide a runnable Docker environment for each instance. Inspired by the multi-layered environment construction approach used in NoCodeBench\cite{deng2025nocode}, we adopt a two-level construction strategy to improve efficiency.

We first build a shared base Docker image for instances originating from the same repository. The base image includes the required system dependencies and standardized runtime configurations, such as Python and CUDA Toolkit versions. With this image as the base image, we further construct instance-specific environments that contain the exact code version and dependencies required for running the corresponding tests.

More specifically, we initially attempt to build the environment using the repository-specific build commands. However, due to issues such as ambiguous dependency specifications or incompatible library updates, some instances fail to build. For example, newer versions of certain libraries may introduce APIs that are incompatible with previously supported configurations or older hardware, requiring specific files to be switched to compatible versions.

To address these instance-specific issues, we record the full build logs during the environment setup process and manually resolve installation errors when they occur. The corresponding fixes are then documented as executable shell commands, such as specifying dependency versions or patching configuration files. These commands are integrated into the instance-specific build scripts to ensure that the environment can be reproduced automatically.

This stage guarantees that each instance is equipped with an independent and fully functional containerized environment, providing a reliable execution foundation for subsequent testing and evaluation.

\begin{table}[t]
\centering
\caption{Distribution of task categories in RealisticTritonBench}
\label{tab:task_types}
\resizebox{\linewidth}{!}{
\begin{tabular}{lcccc}
\toprule
 & Optimization & Modification & New Kernel \\
\midrule
Percentage & 41.93\% & 22.58\% & 35.48\% \\
\bottomrule
\end{tabular}
}
\end{table}

\subsubsection{Refine on Execution Feedback}

After collecting candidate instances and constructing executable environments, we further verify the validity of the tests associated with each instance and perform additional filtering and refinement. Specifically, we first execute all Unit Tests provided for each instance and remove those that fail to pass all Unit Tests, excluding failures caused by hardware-related issues. Since Unit Tests reflect the expected behavior of the modified Triton kernels, instances that cannot pass these tests are considered invalid as they fail to correctly reproduce the behavior of the original PR, and are therefore excluded from the final dataset.

Next, we run the Model Accuracy Test and Latency Test and adjust the corresponding execution commands when necessary based on the observed results. This step is required because the repositories continue to evolve over time, and the interfaces or arguments of testing scripts may change accordingly. As a result, some original test commands may no longer execute correctly in the reconstructed environment. We therefore refine the testing commands according to the actual execution results to ensure that the evaluation scripts remain valid and can reliably measure model accuracy and system latency.

Through these phases, we ultimately obtained 31 task instances, which are grouped into three categories, as summarized in Table~\ref{tab:task_types}.

\section{EXPERIMENT}

\begin{table}[t]
\centering
\caption{Studied Large Language Models}
\label{tab:models}
\resizebox{\linewidth}{!}{
\begin{tabular}{lccc}
\hline
Model & Type & Size & Time \\
\hline
Deepseek-V3.2(non-reasoning) & non-reasoning & 685B & 2025-12-01 \\
Qwen3.5-397B-A17B & reasoning  & 397B & 2026-03-09  \\
Deepseek-V3.2(reasoning) & reasoning & 685B & 2025-12-01 \\
GPT-5.4 & reasoning & unpublished & 2026-03-05 \\
Gemini-3.1 Pro Preview & reasoning & unpublished & 2026-02-19 \\
\hline
\end{tabular}
}
\end{table}

In this section, we evaluate the performance of the state-of-the-art (SOTA) LLMs in different settings and analyze the results. Specifically, we aim
to address the following three research questions.
\begin{itemize}[left=0pt, topsep=0em]
    \item \textbf{RQ1:} How do state-of-the-art LLM-based agents perform on RealisticTritonBench?
    \item \textbf{RQ2:} How do LLMs' performance vary across different task categories?
    \item \textbf{RQ3:} What are the main reasons for the failures of LLMs on RealisticTritonBench?
\end{itemize}

\subsection{Methodology}

\begin{table*}[t]
\centering
\caption{Performance comparison across different LLMs on RealisticTritonBench.}
\label{tab:main_results}
\begin{adjustbox}{max width=0.8\textwidth}
\begin{tabular}{lcccccccc}
\toprule
\textbf{Model} & \textbf{Success (\%)} & \textbf{Applied (\%)} & \textbf{FTP (\%)} & \textbf{UTP (\%)} & \textbf{NR(\%)} & $\mathbf{S_{TTFT}}$ & \textbf{$\mathbf{S_{TPOT}}$} & \textbf{Cost} \\
\midrule

\multicolumn{9}{l}{\textbf{Open-source Models}} \\
\midrule
\textbf{Deepseek-V3.2 (non-reasoning)} & 12.90\% & 93.55\% & 45.16\% & 56.55\% & 45.45\% & 0.9708 & 0.9587 & 17.54M \\
\textbf{Deepseek-V3.2 (reasoning)} & 19.35\% & 96.77\% & 41.94\% & 66.84\% & 40.00\% & 0.9903 & 0.9881 & 25.89M \\
\textbf{Qwen3.5-397B-A17B} & \textbf{25.81\%} & 96.77\% & 45.16\% & 58.12\% & \textbf{58.33\%} & 1.0110 & \textbf{0.9959} & 11.75M \\

\midrule
\multicolumn{9}{l}{\textbf{Closed-source Models}} \\
\midrule
\textbf{GPT-5.4} & 16.13\% & \textbf{100.0\%} & 35.48\% & 52.16\% & 44.44\% & \textbf{1.375} & 0.8948 & \textbf{9.23M} \\
\textbf{Gemini-3.1-Pro-Preview} & 19.35\% & \textbf{100.0\%} & \textbf{48.39\%} & \textbf{67.98\%} & 50.00\% & 0.9424 & 0.9687 & 10.47M \\

\midrule
\midrule
\textbf{Average} & 18.71\% & 97.42\% & 43.23\% & 60.33\% & 47.65\% & 1.0579 & 0.9612 & 14.98M \\
\bottomrule
\end{tabular}
\end{adjustbox}
\end{table*}

\subsubsection{Model Selection}

The tasks in RealisticTritonBench require models to generate complete Triton kernel implementations based on task descriptions and repository context. The process demands strong capabilities in task understanding, Triton kernel generation, and long-context reasoning. Therefore, we select a set of state-of-the-art LLMs for evaluation, as shown in Table~\ref{tab:models}. 
\hjj{Among them, three models are open-source models, including DeepSeek-V3.2 (non-reasoning)\cite{liu2024deepseek}, DeepSeek-V3.2 (reasoning)\cite{liu2024deepseek} and Qwen3.5-397B-A17B\cite{qwen3.5}. The other two are closed-source models, including Gemini-3.1 Pro Preview\cite{google_gemini_31_pro_preview} and GPT-5.4\cite{openai_gpt54_model}.}
These models represent some of the most advanced large language models currently available, with strong performance on code generation and reasoning tasks. 
They differ in model architecture, training paradigms, and functional capabilities, which enables us to investigate how SOTA models with different architectures perform on Triton kernel generation tasks.
\hjj{For models with reasoning capabilities, we set the reasoning effort to high when
configurable and otherwise simply enable thinking.}

\subsubsection{Scaffold Selection}

Finishing Triton kernel generation tasks requires models to retrieve relevant code context from the repository as essential dependency information for implementation. Due to the lack of accurate context, directly prompting an LLM often fails to produce correct and complete implementations. Therefore, we select \textbf{mini-SWE-agent} \cite{yang2024sweagent} as the evaluation scaffold.

\textbf{mini-SWE-agent} is a lightweight implementation derived from the SWE-agent system developed by the Princeton and Stanford teams, designed to simplify the coding-agent pipeline while maintaining strong performance. On the SWE-Bench (Verified) leaderboard, it achieves over 76\% pass rate. Although several agents rank higher, including \textit{live-swe-agent}\cite{xia2025live}, \textit{Sonar Foundation Agent}, \textit{TRAE}\cite{gao2025trae}, \textit{Atlassian Rovo Dev}, and \textit{EPAM AI/Run Developer Agent}, they are not suitable for our evaluation for the following reasons.

First, \textit{Sonar Foundation Agent}, \textit{Atlassian Rovo Dev}, and \textit{EPAM AI/Run Developer Agent} are closed-source systems, making it impossible to obtain their implementations for experimentation. Second, \textit{TRAE} is inactively maintained and currently does not support models without tool-calling capabilities. Finally, although \textit{live-swe-agent} is SOTA and open-source, it modifies the original scaffold through tool evolution. Since our goal is to evaluate the intrinsic capability of models, we follow SWE-Bench and adopt a bash-only tool interface for the scaffold. Taking into account of these factors, mini-SWE-agent is ultimately selected as the scaffold used in our evaluation.

\subsubsection{Evaluation Metrics}

Different from kernel-oriented benchmarks \cite{ouyangkernelbench, li2025tritonbenchbenchmarkinglargelanguage, xing2026flashinfer}, which mainly focus on kernel-level evaluation, we assess the practical performance of generated Triton kernels within the overall framework. We introduce four evaluation metrics to assess the correctness, model numerical stability, and system-level performance of the generated kernels:

\begin{itemize}[left=0pt, topsep=0em]
    \item \textbf{Full Test Pass Rate (FTP\%):} 
    The percentage of tasks whose generated implementations pass all unit tests.

    \item \textbf{Unit Test Pass Rate (UTP\%):} 
    The percentage of unit tests passed out of the total number of unit tests for a task. 

    \item \textbf{Numerical Robustness for Model (NR, T/F):} 
    Evaluates whether the kernel modification introduces numerical instability at the model level. 
    After integrating the generated kernel into the framework, the model is evaluated on a common benchmark. Following the criteria derived from real-world open-source PRs, if the task performance does not degrade, the metric is marked as \textbf{T}; otherwise, it is marked as \textbf{F}.

    \item \textbf{End-to-End Acceleration Speedup:} 
    Measures the improvement in end-to-end latency after deploying the modified kernel in the AI framework compared with the reference implementation under the same workload. 
    We report the speedup of both Time to First Token (TTFT) and Time per Output Token (TPOT), defined as
    \[
    S_{\text{TTFT}}=\frac{\mathrm{TTFT}_{\text{base}}}{\mathrm{TTFT}_{\text{new}}}, 
    \quad
    S_{\text{TPOT}}=\frac{\mathrm{TPOT}_{\text{base}}}{\mathrm{TPOT}_{\text{new}}}.
    \]
    A value greater than 1 indicates a latency reduction compared with the baseline. 
    (We report the average NR and speedup over tasks that pass all unit tests)
\end{itemize}

In addition, following prior work~\cite{zan2025multi, deng2025nocode, jimenez2024swebench}, we also report the success rate of kernel task(\textbf{Success\%}) and \hjj{token cost(\textbf{Cost}) for each model}, the success rate of patch application(\textbf{Applied\%}) as aggregate evaluation metrics. 
A task is considered successful only if all of the following conditions are satisfied:

\begin{itemize}[left=0pt,topsep=0em]
    \item The Unit Test Pass Rate (UTP) matches that of the gold patch implementation.
    \item The Numerical Robustness check passes ($NR = T$).
    \item The end-to-end latency satisfies $S_{\text{TTFT}} \geq 0.98$ and $S_{\text{TPOT}} \geq 0.98$.
\end{itemize}

The first condition ensures functional correctness, the second verifies numerical stability at the model level, and the third ensures that the generated kernel does not significantly degrade system-level inference performance. 
\hjj{
We adopt a tolerance threshold of $0.98$ for latency speedup to account for minor runtime fluctuations, and discuss its sensitivity in our \onlineappendix~\cite{huang2026realistictritonbench}.
}
All experiments are conducted on a server equipped with 8$\times$ NVIDIA RTX 3090 GPUs and averaged over three runs, \hjj{the average variation is 1.08\% for $S_{\text{TTFT}}$ and 0.98\% for $S_{\text{TPOT}}$, indicating stable speedup measurements.}

\hjj{Notably, we adopt the reference implementation from the gold patch as the baseline for two reasons.    
First, these PRs from widely used frameworks such as vLLM underwent multiple rounds of review by experienced kernel experts, making them a credible reference grounded in real-world development practice. 
Second, it is the only available reference implementation in the target framework at that time, making it a meaningful baseline for assessing whether generated patches reach the accepted real-world solution.}

\subsection{RQ1: Performance on RealisticTritonBench}

We evaluate five representative LLMs on RealisticTritonBench, and the results are presented in Table~\ref{tab:main_results}. 
Our findings show that under stricter and more realistic multi-dimensional evaluation criteria, even state-of-the-art LLMs exhibit limited performance on real-world Triton kernel tasks. 
Specifically, Qwen3.5-397B-A17B achieves the best performance, with a task success rate of 25.81\%, indicating that only a small portion of tasks can be fully solved under realistic constraints.

In terms of unit test performance, the average Unit Test Pass Rate (UTP) across all models reaches 60.33\%, and the Full Test Pass Rate (FTP) is 43.23\%. However, the overall task success rate is only 18.71\%, which is significantly lower than the FTP. This notable gap indicates that although models can often generate implementations that pass all unit tests, these implementations may still degrade model accuracy or increase end-to-end latency when deployed in real systems. In other words, passing unit tests alone is insufficient to guarantee the practical usability of generated kernels. Regarding numerical robustness, only 47.65\% of the cases maintain model accuracy without degradation after replacing the original code. This further highlights that correct implementations may also \hjj{introduce} subtle numerical issues that negatively affect model performance. For end-to-end latency, most models achieve an average $\mathbf{S_{TTFT}}$ of around 1.0579, indicating no clear performance improvement compared to the baseline. Although GPT-5.4 achieves a higher $\mathbf{S_{TTFT}}$ of 1.375, this is mainly due to the limited number of correct cases included in its latency evaluation, making the result less representative. Similarly, the average $\mathbf{S_{TPOT}}$ of 0.9612 further suggests that replacing kernels often results in comparable or slightly degraded performance. This observation indicates that performance optimization at the kernel level is still challenging for current LLMs, especially when considering system-level interactions.

Overall, these results demonstrate that RealisticTritonBench effectively exposes the gap between unit test correctness and real system performance, preventing models from concentrating on unit tests alone. By incorporating system-level accuracy and latency evaluation, our benchmark provides a more comprehensive, realistic, and challenging testbed for Triton kernel generation, and better reflects the requirements of real-world deployment scenarios.

\begin{tcolorbox}[colback=gray!15, colframe=gray!50, boxrule=1pt, boxsep=0pt]
\textbf{Summary:} The state-of-the-art LLMs perform poorly on RealisticTritonBench, struggling to generate Triton kernels that simultaneously pass all unit tests, preserve model accuracy, and improve end-to-end latency. This highlights the difficulty of achieving system-level correctness and efficiency, and demonstrates that RealisticTritonBench effectively exposes the gap between unit test success and real-world deployment requirements.
\end{tcolorbox}

\subsection{RQ2: Category-wise Performance Analysis}

Prior work mainly focuses on translating PyTorch kernels into Triton kernels, which corresponds to the New-kernel category in our work. Beyond this setting, we further introduce two additional task categories: Optimization and Modification. The Optimization category targets performance improvements of existing Triton kernels, while the Modification category involves fixing bugs or extending functionality (e.g., adding support for new data types) in existing kernels. In this section, we analyze the experimental results from a category-wise perspective. Table~\ref{tab:category_results} presents the performance of different models across the three task types: Optimization, Modification, and New-kernel.

\noindent\textbf{Optimization.} Models achieve average 23.08\% success rates on Optimization tasks, with relatively high UTP and FTP across all models. This indicates that LLMs are generally capable of generating optimized Triton kernels that preserve the original functionality based on existing implementations. However, the end-to-end speedup remains limited, with most $\mathbf{S_{TTFT}}$ and $\mathbf{S_{TPOT}}$ values close to 1.0, suggesting that models struggle to deliver meaningful performance improvements. Furthermore, numerical robustness remains a challenge, with only 56.19\% of the generated kernels maintaining model accuracy without noticeable degradation. This indicates that even when functional correctness is preserved, the generated implementations may still introduce subtle numerical inconsistencies that affect downstream model behavior. These results highlight the difficulty of \hjj{considering} low-level performance optimization, numerical stability, and system-level efficiency simultaneously.

\noindent\textbf{Modification.} For Modification tasks, models achieve performance comparable to Optimization tasks in terms of UTP and FTP. Both task types share a common characteristic in that they require generating kernels based on existing Triton implementations, while Optimization focuses on improving performance; Modification aims to alter the original functionality, such as adding new features or fixing bugs. The comparable performance results on UTP and FTP further prove that, given existing Triton kernels as references, LLMs are generally capable of generating functionally equivalent implementations. However, numerical robustness remains a \hjj{significant} challenge in this setting. On average, only 20.00\% of the generated kernels maintain model accuracy without degradation, with several models achieving 0\% NR. This indicates that Modification tasks are more likely to introduce subtle numerical issues that degrade model accuracy. It also \hjj{suggests} that while LLMs can follow functional requirements, they lack a deep understanding of numerical stability and edge-case behaviors when modifying existing kernels.

\noindent\textbf{New-kernel.} The average task success rate of only 5.455\% in Table~\ref{tab:category_results} indicates that New-kernel is the most challenging category for current models. Compared with the other two task types, this setting requires the model to implement a functionally equivalent Triton kernel without any prior Triton implementation as reference. Across the 11 New-kernel tasks, on average, only 20\% of the generated kernels are able to pass all unit tests, which is the lowest among all categories. Moreover, even among the limited set of correct kernels, the end-to-end latency metrics are generally below 1.0, and the average numerical robustness (NR) is only 40\%, suggesting that newly generated kernels often fail to match or surpass the performance of existing implementations. These results highlight the fundamental difficulty for LLMs in generating high-quality Triton kernels from scratch when no Triton-specific reference implementation is available.

\begin{table*}[t]
\centering
\caption{Category-wise performance of different models on RealisticTritonBench across task types: Optimization, Modification, and New-kernel.  }
\label{tab:category_results}
\begin{adjustbox}{max width=0.8\textwidth}
\begin{tabular}{llccccccc}
\toprule
\textbf{Task Type} & \textbf{Model} & \textbf{Success (\%)} & \textbf{Applied (\%)} & \textbf{FTP (\%)} &\textbf{UTP (\%)} & \textbf{NR(\%)} & \textbf{$\mathbf{S_{TTFT}}$} & \textbf{$\mathbf{S_{TPOT}}$} \\
\midrule

\multicolumn{9}{l}{\textbf{Optimization}} \\
\midrule
 & Deepseek-V3.2 (non-reasoning) & 15.38\% & \textbf{100.0\%} & 53.85\% & 75.26\% & \textbf{71.43\%} & 0.9442 & 0.9510 \\
 & Deepseek-V3.2 (reasoning) & 7.692\% & \textbf{100.0\%} & 53.85\% & 76.47\% & 42.86\% & 0.9891 & 0.9794 \\
 & Qwen3.5-397B-A17B & \textbf{38.46\%} & \textbf{100.0\%} & \textbf{69.23\%} & \textbf{82.88\%} & 66.67\% & 1.018 & 0.9920 \\
 & GPT-5.4 & 23.08\% & \textbf{100.0\%} & 46.15\% & 53.65\% & 50.00\% & \textbf{1.812} & 0.9001 \\
 & Gemini-3.1-Pro-Preview & 30.77\% & \textbf{100.0\%} & 61.54\% & 75.61\% & 50.00\% & 0.9964 & \textbf{1.005} \\
 & \textbf{Average} & 23.08\% & \textbf{100.0\%} & 56.92\% & 72.78\% & 56.19\% & 1.152 & 0.9654 \\

\midrule
\multicolumn{9}{l}{\textbf{Modification}} \\
\midrule
 & Deepseek-V3.2 (non-reasoning) & 28.57\% & \textbf{100.0\%} & \textbf{71.43\%} & 71.43\% & 0.000\% & 1.027 & 0.9824 \\
 & Deepseek-V3.2 (reasoning) & \textbf{42.86\%} & \textbf{100.0\%} & 57.14\% & \textbf{79.92\%} & 0.000\% & 1.002 & 1.008 \\
 & Qwen3.5-397B-A17B & \textbf{42.86\%} & \textbf{100.0\%} & 57.14\% & 62.34\% & \textbf{50.00\%} & 0.9825 & 1.004 \\
 & GPT-5.4 & 14.29\% & \textbf{100.0\%} & 28.57\% & 54.10\% & 0.000\% & 1.026 & 1.019 \\
 & Gemini-3.1-Pro-Preview & 28.57\% & \textbf{100.0\%} & 57.14\% & 73.19\% & \textbf{50.00\%} & \textbf{1.033} & \textbf{1.024} \\
 & \textbf{Average} & 31.43\% & \textbf{100.0\%} & 54.29\% & 68.20\% & 20.00\% & 1.014 & 1.007 \\

\midrule
\multicolumn{9}{l}{\textbf{New-kernel}} \\
\midrule
 & Deepseek-V3.2 (non-reasoning) & 0.000\% & 81.81\% & 18.18\% & 24.97\% & 0.000\% & 0.9791 & 0.9500 \\
 & Deepseek-V3.2 (reasoning) & \textbf{18.18\%} & 90.90\% & 18.18\% & 47.14\% & \textbf{100.0\%} & 0.9825 & 0.9990 \\
 & Qwen3.5-397B-A17B & 0.000\% & 90.90\% & 9.091\% & 26.16\% & 0.000\% & \textbf{1.007} & \textbf{1.014} \\
 & GPT-5.4 & 9.091\% & \textbf{100.0\%} & \textbf{27.27\%} & 49.17\% & 50.00\% & 0.8053 & 0.8559 \\
 & Gemini-3.1-Pro-Preview & 0.000\% & \textbf{100.0\%} & \textbf{27.27\%} & \textbf{55.65\%} & 50.00\% & 0.7380 & 0.8359 \\
 & \textbf{Average}  & 5.455\% & \hjj{\textbf{93.94\%}} & 20.00\% & 40.62\% & 40.00\% & 0.9025 & 0.9310 \\

\bottomrule
\end{tabular}
\end{adjustbox}
\end{table*}

\begin{tcolorbox}[colback=gray!15, colframe=gray!50, boxrule=1pt, boxsep=0pt, before skip=10pt, after skip=0pt]
\noindent\textbf{Summary:} LLMs perform well in functional correctness when modifying existing Triton kernels but struggle with performance optimization and numerical robustness. When no Triton-based implementation is provided, performance drops significantly, revealing limited capability in generating high-quality kernels from scratch.
\end{tcolorbox}

\subsection{RQ3: Failure Analysis }

To better understand the limitations of LLMs in Triton kernel generation, we manually inspect the failed cases and identify the key reasons that lead to their poor performance on RealisticTritonBench. Through this analysis, we reveal the limitations of current LLMs in real-world Triton kernel generation tasks.

\begin{figure}[t]
\centering
\begin{minipage}{1.0\linewidth}
\begin{minted}[
fontsize=\small,
bgcolor=gray!10,
linenos=false,
breaklines=true,
breakanywhere=true,
breaksymbolleft=,
breaksymbolright=,
breakindent=2em 
]{python}
# Target Function
@triton.jit
def kernel_unified_attention_2d(...):
    ...
    cur_batch_in_all_start_index = tl.load(query_start_len_ptr + seq_idx)
    cur_batch_in_all_stop_index = tl.load(query_start_len_ptr + seq_idx + 1)
    cur_batch_query_len = cur_batch_in_all_stop_index - cur_batch_in_all_start_index
    ...
    seq_len = tl.load(seq_lens_ptr + seq_idx)
    ...
    q_end_pos = tl.min(q_end_pos, cur_batch_query_len)
\end{minted}
\end{minipage}
\caption{An example of incorrect usage of the Triton API.}
\label{fig:case1}
\end{figure}

\noindent\textbf{Insufficient fundamental capability in Triton programming}. 
On average, 56.77\% of the cases fail to pass all unit tests, indicating that generating fully correct Triton kernels remains highly challenging for current LLMs. A fundamental prerequisite for completing these tasks is the ability to produce compilable and executable code that strictly adheres to Triton’s programming constraints, including its type system, supported control flow, and memory access semantics. However, our analysis of failure cases shows that a substantial portion of errors originate from violations of these low-level constraints. Among the cases that fail to pass all unit tests, \textbf{64.71\%} fall into this category. Concretely, models frequently hallucinate non-existent APIs (e.g. \texttt{tl.info}, \texttt{tl.finfo}), misuse existing primitives (e.g. incorrect usage of \texttt{tl.min}), or produce code that violates Triton’s strict typing rules. In addition, several cases exhibit incorrect memory access patterns, such as mismatched tensor shapes or invalid pointer arithmetic in \texttt{tl.load}, which result in runtime errors or silent incorrect behavior. There are also instances of unsupported control flow constructs (e.g. \texttt{break}) and missing essential components (e.g. decorators), further preventing successful execution. These results indicate that current LLMs lack a systematic and reliable understanding of Triton’s low-level programming model. In particular, they struggle to internalize the strict constraints imposed by the compiler and runtime, including type safety, valid API usage, and explicit memory layout requirements.

For example, as illustrated in Figure~\ref{fig:case1}, the model incorrectly uses the \texttt{tl.min} API. In \texttt{kernel\_unified\_attention\_2d}, both \texttt{cur\_batch\_query\_len} and \texttt{q\_end\_pos} are scalar values (i.e., 0-D Triton tensors). However, \texttt{tl.min} in Triton is designed as a reduction API, expecting an input tensor together with a reduction axis, rather than two scalar operands for binary comparison. Therefore, \texttt{tl.min(q\_end\_pos, cur\_batch\_query\_len)} does not match the intended semantics of taking the minimum of two scalar values. The correct API here should be \texttt{tl.minimum}, which performs an elementwise minimum between two operands.

\noindent\textbf{Lack of understanding of kernel semantics in real-world repository code}. 
In addition to violations of low-level programming constraints, we observe a \textbf{29.41\%} of failures that stem from incorrect implementations of kernel semantics in concrete code repositories. Successfully completing these tasks requires the model to comprehensively reason about kernel behavior in repositories, including boundary conditions, corner cases, and fine-grained implementation details that are critical to correctness and numerical behavior. We observe that LLMs frequently make mistakes in implementation details. As illustrated in Figure~\ref{fig:case2}, the generated implementation fails to handle segment processing. Instead of returning early for invalid segments, it incorrectly writes back incorrect \texttt{M\_seg}, \texttt{L\_seg}, and \texttt{acc\_seg}, thereby corrupting the subsequent segment-level reduction. This case further highlights that LLMs lack a comprehensive understanding of kernel semantics in real-world implementations. In real-world kernel generation tasks, correct implementations require careful consideration of both the task objectives and the surrounding code context to properly handle different scenarios, including boundary conditions and edge cases. LLMs frequently struggle to correctly handle corner cases and often ignore implicit conditions,  ultimately leading to functional inconsistencies or incorrect numerical results. This indicates that LLMs lack a comprehensive understanding of kernel semantics in real-world repository contexts.

\begin{figure}[t]
\centering
\begin{minipage}{1.0\linewidth}
\begin{minted}[
fontsize=\small,
bgcolor=gray!10,
linenos=false,
breaklines=true,
breakanywhere=true,
breaksymbolleft=,
breaksymbolright=,
breakindent=2em
]{python}
# Reference Implementation
def kernel_unified_attention_3d(...):
    ...
    # Reference Implementation
    blocks_per_segment = cdiv_fn(seq_len, NUM_SEGMENTS_PER_SEQ * BLOCK_SIZE)
    if segm_idx*blocks_per_segment*BLOCK_SIZE >= seq_len:
        return
    # Generated implementation
    segment_len = cdiv_fn(num_blocks, NUM_SEGMENTS_PER_SEQ)
    start_block = segment_idx * segment_len
    end_block = tl.minimum(start_block + segment_len,...)
        ...
    tl.store(segm_max_ptr + segm_offset, M, ...)
    tl.store(segm_expsum_ptr + segm_offset, L, ...)
\end{minted}
\end{minipage}
\caption{An example where the generated implementation incorrectly writes back invalid segments instead of skipping}
\label{fig:case2}
\end{figure}

\textbf{Lack of attention to performance and numerical stability during implementation}.
In our experiments, models successfully generate Triton kernels that pass all correctness checks in 43.23\% of cases. 
However, in 25.16\% of cases, replacing the original implementation with the generated kernel leads to degraded model accuracy or end-to-end performance. 
Through case analysis, we find that generated kernels often introduce unnecessary operations or miss critical optimizations compared to reference kernels. 

For example, as shown in Figure~\ref{fig:case3}, the gold implementation encloses the penalty-related computation within an \texttt{if use\_penalty} branch, thereby avoiding redundant arithmetic operations and predicate computations. 
In contrast, the generated implementation executes these computations unconditionally, resulting in increased latency. 
Furthermore, we observe that generated kernels may degrade model accuracy due to subtle numerical issues introduced during implementation. 
For example, changes to the specialization strategy (e.g., conversion to \texttt{tl.constexpr} and modifying \texttt{do\_not\_specialize}) may affect Triton’s execution behavior, leading to small numerical deviations that accumulate and impact model accuracy. These cases show that LLMs lack of enough attention to performance and numerical stability in kernel implementation.

\begin{figure}[t]
\centering
\begin{minipage}{1.0\linewidth}
\begin{minted}[
fontsize=\small,
bgcolor=gray!10,
linenos=false,
breaklines=true,
breakanywhere=true,
breaksymbolleft=,
breaksymbolright=,
breakindent=2em
]{python}
def _penalties_and_temperature_kernel(...):
    ...
    # Reference Implementation
    if use_penalty:
        req_state_idx = tl.load(idx_mapping_ptr + batch_idx)
        output_bin_counts = tl.load(output_bin_counts_ptr + req_state_idx * output_bin_counts_stride + block,mask=mask)
        output_bin_mask = output_bin_counts > 0
    # Generated Implementation
    req_state_idx = tl.load(idx_mapping_ptr + batch_idx)
    output_bin_counts = tl.load( output_bin_counts_ptr + req_state_idx * output_bin_counts_stride + block, mask=mask,)
    ...
\end{minted}
\end{minipage}
\caption{An example where the generated implementation performs redundant computation}
\label{fig:case3}
\end{figure}

\begin{tcolorbox}[colback=gray!15, colframe=gray!50, boxrule=1pt, boxsep=0pt]
\noindent\textbf{Summary:} LLM failures in real-world Triton kernel generation stem from three key limitations: insufficient mastery of Triton programming constraints, incomplete understanding of kernel semantics in real-world codebases, and lack of attention to performance efficiency and numerical stability.
\end{tcolorbox}

\section{Discussion}

\subsection{Necessity of Framework-Level Evaluation}

\hjj{Among the failures reported in our evaluation, a portion of them, such as missing boundary masks, could in principle be caught by stronger isolated kernel-level tests. 
However, the failures related to numerical deviations, whose acceptability is decided by downstream computation,  can hardly be detected by kernel-level tests. 
}

\begin{table}[t]
\centering
\caption{Reward-hacking strategies catalogued by SOL-ExecBench~\cite{lin2026sol}}
\label{tab:reward_hacking}
\small
\setlength{\tabcolsep}{4pt}
\begin{tabular}{lp{5.6cm}}
\toprule
\textbf{Category} & \textbf{Exploit Description} \\
\midrule
Concurrency & Offloads computation to background threads, side streams, or forked processes to escape the timed region. \\
\midrule
State \& Caching & Returns cached or lazily materialized results, or behaves correctly only when a check is detected. \\
\midrule
Environment & Monkey-patches timing or comparison utilities, or silently downgrades numerical precision for speed. \\
\bottomrule
\end{tabular}
\end{table}

\hjj{For example, in one case from our evaluation, i.e., \texttt{\_fused\_\allowbreak moe\_\allowbreak lora\_\allowbreak kernel} (shown in Figure~\ref{fig:moe_lora_case}), the generated kernel introduces two numerical deviations. One arises from parallelism in computation: the generated kernel replaces the gold implementation's sequential accumulation over the K dimension with parallel K-splitting merged via \texttt{atomic\_add}. Under this parallel strategy, the commit order of the partial sums is decided by GPU scheduling races. Since floating-point addition is not associative, the outputs deviate slightly from the reference and even vary across runs on the same input. The other arises from rounding of partial sums: the gold implementation rounds each value to BF16 before writing it, whereas the generated kernel rounds each value directly to the output element type. This difference in rounding behavior can introduce a small but systematic numerical discrepancy when the output dtype is not BF16. Both numerical deviations are tiny per call, and the kernel passes all unit tests, yet they accumulate across dozens of stacked MoE-LoRA layers and thousands of autoregressive decoding steps, eventually degrading GSM8K accuracy. These failures can hardly be detected by the tolerance threshold of a kernel-level unit test, because it is hard to derive an appropriate threshold for two reasons. First, deriving an acceptable error threshold requires framework-level analysis. Determining how much per-call deviation can be tolerated without degrading task accuracy requires tracing how the error propagates through the actual model's stacked layers, discrete routing decisions, and autoregressive decoding process, none of which is a property of the kernel. 
Second, the acceptable error threshold is deployment-specific rather than kernel-specific. 
Although the same kernel may be reused across models, tasks, and runtime configurations, these contexts determine how errors propagate and accumulate. 
Models may invoke the kernel across more layers, longer decoding sequences allow errors to accumulate over steps, and different serving precisions change the magnitude of each deviation. 
These system-level effects are difficult, if not impossible, to capture with kernel-level unit tests, making framework-level evaluation essential.
}

\begin{figure}[t]
\centering
\begin{minipage}{1.0\linewidth}
\begin{minted}[
fontsize=\small,
bgcolor=gray!10,
linenos=false,
breaklines=true,
breakanywhere=true,
breaksymbolleft=,
breaksymbolright=,
breakindent=2em
]{python}
for k in range(0, grid_k):
    ...
    accumulator += tl.dot(a, b)
    a_ptrs += BLOCK_SIZE_K * SPLIT_K * stride_ak
    b_ptrs += BLOCK_SIZE_K * SPLIT_K * stride_bk
...
# Reference Implementation
accumulator = accumulator.to(tl.bfloat16)  # fixed to bf16
tl.store(c_ptrs, accumulator, mask=c_mask)                          
# Generated Implementation
tl.atomic_add(c_ptrs, accumulator.to(c_ptrs.dtype.element_ty), mask=c_mask) # rounds to output dtype, not bf16
...
\end{minted}
\end{minipage}
\caption{An example of numerical deviations whose acceptability is decided by downstream computation}
\label{fig:moe_lora_case}
\end{figure}

\subsection{Reward Hacking Mitigation}

\hjj{Reward hacking occurs when a model exploits loopholes in the evaluation environment to maximize its score without genuinely solving the underlying task\cite{lin2026sol}. Prior work\cite{lin2026sol} systematically categorizes hacking strategies in kernel generation into three groups: concurrency-based hacks, state-and-caching hacks, and environment manipulation, as summarized in Table \ref{tab:reward_hacking}. These hacks arise from a fundamental limitation of kernel-level testing: it cannot fully capture the correctness and performance of a kernel as it operates within the complete framework. Although kernel-level metrics are often strongly correlated with the metrics that matter in deployment, they are not equivalent.  To mitigate the gap, RealisticTritonBench evaluates each kernel in a setting that closely mirrors its real-world usage. Specifically, (1) the generated kernel is integrated through the same framework call paths used in deployment; (2) it is validated using the same benchmarks that framework developers use to test kernels before deployment; (3) performance is measured using the end-to-end serving latency, i.e., time-to-first-token ($S_{\text{TTFT}}$) and time-per-output-token ($S_{\text{TPOT}}$), and model accuracy, which actual deployment cares about.
}

\begin{figure}[t]
\centering
\begin{minipage}{1.0\linewidth}
\begin{minted}[
fontsize=\small,
bgcolor=gray!10,
linenos=false,
breaklines=true,
breakanywhere=true,
breaksymbolleft=,
breaksymbolright=,
breakindent=2em
]{python}
# Hacked kernel: escape the timed stream
def hacked_kernel(x):
    out = torch.empty_like(x)
    with torch.cuda.stream(side_stream):
        real_kernel[grid](x, out, ...)
    return out  
# Kernel-level harness (simplified)
start.record()        # on default stream
hacked_kernel(x)
end.record()          # on default stream
torch.cuda.synchronize()
t = start.elapsed_time(end)  # launch overhead only
assert_close(hacked_kernel(x), ref)  # passes
\end{minted}
\end{minipage}
\caption{A stream-injection hack that defeats kernel-level timing checks.}
\label{fig:stream_hack}
\end{figure}

\hjj{This design mitigates all three categories of reward hacking listed in Table \ref{tab:reward_hacking}. 
First, concurrency-based hacks are ineffective because an external client measures TTFT and TPOT as the wall-clock time from request submission to response delivery. 
This interval encompasses work performed by any thread, CUDA stream, or process. Consequently, hidden work in concurrency-based hacks is either included in the measured latency or results in incorrect outputs. 
Second, state-and-caching hacks are unlikely to succeed because kernel outputs are immediately consumed by downstream computations across thousands of invocations under evolving runtime states. Cached, stale, or lazily materialized results therefore quickly cause numerical divergence and fail the model-accuracy evaluation. 
Third, environment-manipulation attacks are neutralized because the timing client and accuracy harness run in separate processes beyond the reach of the generated kernel. 
The kernel therefore cannot improve its score by modifying the evaluation utilities. Similarly, silently reducing numerical precision for additional speed is ineffective when the resulting errors propagate through the model and degrade its end-to-end accuracy. 

Figure~\ref{fig:stream_hack} illustrates a stream-injection attack. 
The hacked wrapper launches the actual computation on a separate CUDA stream and returns immediately. 
An in-process timer that synchronizes only the default stream consequently measures little more than the launch overhead, while a later correctness check may still pass after the delayed computation has completed. 
This strategy is ineffective in RealisticTritonBench because the external client continues measuring time until it receives the final response. 
Thus, computation hidden on a separate stream is either fully reflected in TTFT and TPOT or causes incorrect outputs that fail the accuracy evaluation.
}

\section{Threats To Validity}

\hjj{
The first threat arises from the automatic generation of task descriptions. 
In our construction pipeline, the task description of each instance is generated by an LLM from the corresponding pull request (PR) information and code changes. The LLM may misinterpret the developer's intent, or omit implicit implementation details. 
To mitigate this, we manually review all generated task descriptions and refine them when necessary to ensure they accurately reflect the intent of the original PR.
The second threat concerns the generalization of our results across models. 
We evaluate only a set of representative LLMs, so the conclusions may depend on the selected models and may not fully generalize to all existing models. 
To alleviate this, we select mainstream models from different model families with varying capabilities to improve the representativeness of the evaluation.
The third threat stems from the evaluation scaffold. 
Main experiments are conducted with mini-SWE-agent, while different scaffolds may adopt different retrieval strategies and execution workflows that could influence model performance. 
We choose mini-SWE-agent because it represents the state-of-the-art level among open-source coding agents and is also the official agent adopted by SWE-bench for evaluating model performance.
Furthermore, we additionally evaluate GPT-5.4 with its native scaffold Codex to examine the influence of the scaffold choice and provide the results in our \onlineappendix~\cite{huang2026realistictritonbench}.
The last threat is potential data contamination. 
Our benchmark is constructed from publicly available repositories, so the evaluated models may have been exposed to the relevant code during training. We analyze this risk and provide results in our \onlineappendix~\cite{huang2026realistictritonbench}, which suggest that contamination is unlikely to affect our conclusions.
}

\section{RELATED WORK}

\subsection{Triton Kernel Generation Datasets}

\hjj{To evaluate the quality of Triton kernels generated by LLMs, several benchmarks have been proposed.
KernelBench \citep{ouyangkernelbench}, the first benchmark for GPU kernel generation, collects 250 AI workloads of three complexity levels and requires models to generate optimized kernels from PyTorch reference implementations.
TritonBench \citep{li2025tritonbenchbenchmarkinglargelanguage} targets Triton specifically, with tasks curated from highly starred Triton repositories and selected PyTorch implementations.
More recently, FlashInferBench \citep{xing2026flashinfer} extends the pipeline to integration within practical AI frameworks, enabling end-to-end evaluation.}

\hjj{Despite these efforts, the first two benchmarks formulate tasks as PyTorch-to-Triton translation and evaluate kernels in isolation with kernel-level metrics. These kernel-level benchmarks do not capture real-world development scenarios such as performance optimization.  FlashInferBench integrates generated kernels into practical frameworks, but still differs from our work in two aspects. 
(1) \textbf{Task formulation.} It targets generating new kernels for a set of predefined kernel definitions, whereas RealisticTritonBench derives tasks from merged PRs in popular AI frameworks, covering Optimization, Modification, and New Kernel. 
(2) \textbf{Evaluation metrics.} It mainly relies on the kernel-level \texttt{fast\_p} metric complemented with request-level latency, whereas RealisticTritonBench evaluates unit-test correctness, model numerical robustness, and end-to-end TTFT/TPOT latency, directly measuring the impact on accuracy and serving performance after deployment.
}

\subsection{LLM for Automated Triton Generation}
\hjj{
LLMs have been widely applied to general code generation tasks \cite{jiang2025agentic, ma2024lingma, xia2024agentless, wang2025aegis, ruan2025specrover, ma2025alibaba, jiang2025issue, wang2026icore} and have also shown promising potential for Triton kernel generation.
Existing approaches can be categorized into three groups: Domain Model Training, Agent-Based Pipelines, and Agentic RL Training.

Domain Model Training fine-tunes LLMs on systematically collected or synthesized Triton-specific data. KernelLLM \citep{fisches2025kernelllm} performs instruction tuning on compiler-aligned PyTorch-Triton pairs, AutoTriton \citep{li2025autotritonautomatictritonprogramming} combines supervised fine-tuning with GRPO-based reinforcement learning~\cite{shao2024deepseekmath}, and TritonRL \citep{woo2025tritonrl} further introduces hierarchical reward decomposition.
Agent-Based Pipelines employ LLMs as coding agents that iteratively refine kernels through a generate-execute-refine loop driven by tools and runtime feedback, as exemplified by KernelFalcon \citep{kernelfalcon2025}, TritorX \cite{hammond2026agentic}, and AKG Kernel Agent \cite{du2025akg}.
Agentic RL further optimizes the agent's decision-making via reinforcement learning, e.g., QiMeng-Kernel \citep{zhu2026qimeng} adopts a two-stage paradigm that first learns hardware-aware optimization strategies and then implements them into efficient kernels.
}

\section{CONCLUSION}

In this paper, we introduce RealisticTritonBench, a benchmark for evaluating LLM-based Triton kernel generation under realistic development settings. Unlike prior benchmarks that focus mainly on PyTorch-to-Triton translation and kernel-level evaluation, RealisticTritonBench is constructed from real pull requests collected from open-source AI frameworks and covers diverse development scenarios, including kernel optimization, modification, and new operator implementation. We further design a comprehensive evaluation pipeline that integrates unit testing with end-to-end model accuracy and system-level latency evaluation. Through experiments on several representative LLMs, we provide a systematic analysis of their capabilities and limitations in practical Triton kernel development tasks. Our results show that current LLMs lack the capability to generate correct and performance-efficient Triton kernels in real-world tasks. We envision RealisticTritonBench to facilitate the development of \hjj{Triton} kernel generation methods that are effective under realistic scenarios.

\section{DATA AVAILABILITY}

Our code and data are available at \url{https://doi.org/10.5281/zenodo.19221469} and \url{https://github.com/ZJU-CTAG/RealisticTritonBench}

\begin{acks}
This research is supported by the National Natural Science Foundation of China (No.92582107) and Zhejiang Provincial Natural Science Foundation of China (No.LZ25F020003).
\end{acks}

\bibliographystyle{ACM-Reference-Format}
\bibliography{references}

@article{lin2026sol,
  title={SOL-ExecBench: Speed-of-Light Benchmarking for Real-World GPU Kernels Against Hardware Limits},
  author={Lin, Edward and Modi, Sahil and Hari, Siva Kumar Sastry and Huang, Qijing and Ye, Zhifan and Qin, Nestor and Zhou, Fengzhe and Zhang, Yuan and Wang, Jingquan and Damani, Sana and others},
  journal={arXiv preprint arXiv:2603.19173},
  year={2026}
}

@misc{google_gemini_31_pro_preview,
  title        = {{Gemini 3.1 Pro Preview}},
  author       = {{Google}},
  year         = {2026},
  howpublished = {\url{https://ai.google.dev/gemini-api/docs/models/gemini-3.1-pro-preview}},
  note         = {accessed: 2026-03}
}

@misc{openai_gpt54_model,
  title        = {{GPT-5.4 Model}},
  author       = {{OpenAI}},
  year         = {2026},
  howpublished = {\url{https://developers.openai.com/api/docs/models/gpt-5.4}},
  note         = {Accessed: 2026-03}
}

@misc{qwen3.5,
    title  = {{Qwen3.5}: Towards Native Multimodal Agents},
    author = {{Qwen Team}},
    month  = {February},
    year   = {2026},
    url    = {https://qwen.ai/blog?id=qwen3.5}
}

@article{liu2024deepseek,
  title={Deepseek-v3 technical report},
  author={Liu, Aixin and Feng, Bei and Xue, Bing and Wang, Bingxuan and Wu, Bochao and Lu, Chengda and Zhao, Chenggang and Deng, Chengqi and Zhang, Chenyu and Ruan, Chong and others},
  journal={arXiv preprint arXiv:2412.19437},
  year={2024}
}

@article{shao2024deepseekmath,
  title={Deepseekmath: Pushing the limits of mathematical reasoning in open language models},
  author={Shao, Zhihong and Wang, Peiyi and Zhu, Qihao and Xu, Runxin and Song, Junxiao and Bi, Xiao and Zhang, Haowei and Zhang, Mingchuan and Li, YK and Wu, Yang and others},
  journal={arXiv preprint arXiv:2402.03300},
  year={2024}
}

@inproceedings{ma2025alibaba,
  title={Alibaba lingmaagent: Improving automated issue resolution via comprehensive repository exploration},
  author={Ma, Yingwei and Yang, Qingping and Cao, Rongyu and Li, Binhua and Huang, Fei and Li, Yongbin},
  booktitle={Proceedings of the 33rd ACM International Conference on the Foundations of Software Engineering},
  pages={238--249},
  year={2025}
}

@inproceedings{ruan2025specrover,
  title={Specrover: Code intent extraction via llms},
  author={Ruan, Haifeng and Zhang, Yuntong and Roychoudhury, Abhik},
  booktitle={2025 IEEE/ACM 47th International Conference on Software Engineering (ICSE)},
  pages={963--974},
  year={2025},
  organization={IEEE}
}

@inproceedings{wang2025aegis,
  title={Aegis: An agent-based framework for bug reproduction from issue descriptions},
  author={Wang, Xinchen and Gao, Pengfei and Meng, Xiangxin and Peng, Chao and Hu, Ruida and Lin, Yun and Gao, Cuiyun},
  booktitle={Proceedings of the 33rd ACM International Conference on the Foundations of Software Engineering},
  pages={331--342},
  year={2025}
}

@article{ma2024lingma,
  title={Swe-gpt: A process-centric language model for automated software improvement},
  author={Ma, Yingwei and Cao, Rongyu and Cao, Yongchang and Zhang, Yue and Chen, Jue and Liu, Yibo and Liu, Yuchen and Li, Binhua and Huang, Fei and Li, Yongbin},
  journal={Proceedings of the ACM on Software Engineering},
  volume={2},
  number={ISSTA},
  pages={2362--2383},
  year={2025},
}

@article{xia2024agentless,
  title={Demystifying llm-based software engineering agents},
  author={Xia, Chunqiu Steven and Deng, Yinlin and Dunn, Soren and Zhang, Lingming},
  journal={Proceedings of the ACM on Software Engineering},
  volume={2},
  number={FSE},
  pages={801--824},
  year={2025},
}

@inproceedings{hu2025lightllm,
  title={LightLLM: A versatile large language model for predictive light sensing},
  author={Hu, Jiawei and Jia, Hong and Hassan, Mahbub and Yao, Lina and Kusy, Brano and Hu, Wen},
  booktitle={Proceedings of the 23rd ACM Conference on Embedded Networked Sensor Systems},
  pages={158--171},
  year={2025}
}

@misc{ggerganov2023llamacpp,
  author       = {Georgi Gerganov and contributors},
  title        = {llama.cpp: LLM Inference in C/C++},
  year         = {2023},
  howpublished = {\url{https://github.com/ggml-org/llama.cpp}},
  note         = {GitHub repository}
}

@article{dao2022flashattention,
  title={Flashattention: Fast and memory-efficient exact attention with io-awareness},
  author={Dao, Tri and Fu, Dan and Ermon, Stefano and Rudra, Atri and R{\'e}, Christopher},
  journal={Advances in neural information processing systems},
  volume={35},
  pages={16344--16359},
  year={2022}
}

@misc{wen2025multikernelbenchmultiplatformbenchmarkkernel,
      title={MultiKernelBench: A Multi-Platform Benchmark for Kernel Generation}, 
      author={Zhongzhen Wen and Yinghui Zhang and Zhong Li and Zhongxin Liu and Linna Xie and Tian Zhang},
      journal={arXiv preprint arXiv:2507.17773},
      year={2025},
}

@article{ye2025flashinfer,
  title={Flashinfer: Efficient and customizable attention engine for llm inference serving},
  author={Ye, Zihao and Chen, Lequn and Lai, Ruihang and Lin, Wuwei and Zhang, Yineng and Wang, Stephanie and Chen, Tianqi and Kasikci, Baris and Grover, Vinod and Krishnamurthy, Arvind and others},
  journal={Proceedings of Machine Learning and Systems},
  volume={7},
  year={2025}
}

@inproceedings{ali2020batch,
  title={Batch: Machine learning inference serving on serverless platforms with adaptive batching},
  author={Ali, Ahsan and Pinciroli, Riccardo and Yan, Feng and Smirni, Evgenia},
  booktitle={SC20: International Conference for High Performance Computing, Networking, Storage and Analysis},
  pages={1--15},
  year={2020},
  organization={IEEE}
}

@article{zhang2025efficient,
  title={Efficient Mixed-Precision Large Language Model Inference with TurboMind},
  author={Zhang, Li and Jiang, Youhe and He, Guoliang and Chen, Xin and Lv, Han and Yao, Qian and Fu, Fangcheng and Chen, Kai},
  journal={arXiv preprint arXiv:2508.15601},
  year={2025}
}

@inproceedings{abadi2016tensorflow,
  title={$\{$TensorFlow$\}$: a system for $\{$Large-Scale$\}$ machine learning},
  author={Abadi, Mart{\'\i}n and Barham, Paul and Chen, Jianmin and Chen, Zhifeng and Davis, Andy and Dean, Jeffrey and Devin, Matthieu and Ghemawat, Sanjay and Irving, Geoffrey and Isard, Michael and others},
  booktitle={12th USENIX symposium on operating systems design and implementation (OSDI 16)},
  pages={265--283},
  year={2016}
}

@inproceedings{li2025llava,
  title={Llava-st: A multimodal large language model for fine-grained spatial-temporal understanding},
  author={Li, Hongyu and Chen, Jinyu and Wei, Ziyu and Huang, Shaofei and Hui, Tianrui and Gao, Jialin and Wei, Xiaoming and Liu, Si},
  booktitle={Proceedings of the IEEE/CVF Conference on Computer Vision and Pattern Recognition},
  pages={8592--8603},
  year={2025}
}

@article{shao2025deepseekmath,
  title={Deepseekmath-v2: Towards self-verifiable mathematical reasoning},
  author={Shao, Zhihong and Luo, Yuxiang and Lu, Chengda and Ren, ZZ and Hu, Jiewen and Ye, Tian and Gou, Zhibin and Ma, Shirong and Zhang, Xiaokang},
  journal={arXiv preprint arXiv:2511.22570},
  year={2025}
}

@article{wiedemer2025video,
  title={Video models are zero-shot learners and reasoners},
  author={Wiedemer, Thadd{\"a}us and Li, Yuxuan and Vicol, Paul and Gu, Shixiang Shane and Matarese, Nick and Swersky, Kevin and Kim, Been and Jaini, Priyank and Geirhos, Robert},
  journal={arXiv preprint arXiv:2509.20328},
  year={2025}
}

@article{cao2026qwen3,
  title={Qwen3-Coder-Next Technical Report},
  author={Cao, Ruisheng and Chen, Mouxiang and Chen, Jiawei and Cui, Zeyu and Feng, Yunlong and Hui, Binyuan and Jing, Yuheng and Li, Kaixin and Li, Mingze and Lin, Junyang and others},
  journal={arXiv preprint arXiv:2603.00729},
  year={2026}
}

@article{shoeybi2019megatron,
  title={Megatron-lm: Training multi-billion parameter language models using model parallelism},
  author={Shoeybi, Mohammad and Patwary, Mostofa and Puri, Raul and LeGresley, Patrick and Casper, Jared and Catanzaro, Bryan},
  journal={arXiv preprint arXiv:1909.08053},
  year={2019}
}

@article{hsu2024liger,
  title={Liger kernel: Efficient triton kernels for llm training},
  author={Hsu, Pin-Lun and Dai, Yun and Kothapalli, Vignesh and Song, Qingquan and Tang, Shao and Zhu, Siyu and Shimizu, Steven and Sahni, Shivam and Ning, Haowen and Chen, Yanning},
  journal={arXiv preprint arXiv:2410.10989},
  year={2024}
}

@article{jiang2025agentic,
  title={Agentic Software Issue Resolution with Large Language Models: A Survey},
  author={Jiang, Zhonghao and Lo, David and Liu, Zhongxin},
  journal={arXiv preprint arXiv:2512.22256},
  year={2025}
}

@inproceedings{zan2025multi,
  title={Multi-SWE-bench: A Multilingual Benchmark for Issue Resolving},
  author={Zan, Daoguang and Huang, Zhirong and Liu, Wei and Chen, Hanwu and Xin, Shulin and Zhang, Linhao and Liu, Qi and Li, Aoyan and Chen, Lu and Zhong, Xiaojian and others},
  booktitle={The Thirty-ninth Annual Conference on Neural Information Processing Systems Datasets and Benchmarks Track}
}

@inproceedings{yang2024sweagent,
  title={Swe-agent: Agent-computer interfaces enable automated software engineering},
  author={Yang, John and Jimenez, Carlos E and Wettig, Alexander and Lieret, Kilian and Yao, Shunyu and Narasimhan, Karthik R and Press, Ofir},
  booktitle={The Thirty-eighth Annual Conference on Neural Information Processing Systems},
  year={2024}
}

@inproceedings{jimenez2024swebench,
  title={Swe-bench: Can language models resolve real-world github issues?},
  author={Jimenez, Carlos E and Yang, John and Wettig, Alexander and Yao, Shunyu and Pei, Kexin and Press, Ofir and Narasimhan, Karthik},
  booktitle={International Conference on Learning Representations},
  volume={2024},
  pages={54107--54157},
  year={2024}
}

@article{deng2025nocode,
  title={Nocode-bench: A benchmark for evaluating natural language-driven feature addition},
  author={Deng, Le and Jiang, Zhonghao and Cao, Jialun and Pradel, Michael and Liu, Zhongxin},
  journal={arXiv preprint arXiv:2507.18130},
  year={2025}
}

@article{paszke2019pytorchimperativestylehighperformance,
  title={Pytorch: An imperative style, high-performance deep learning library},
  author={Paszke, Adam and Gross, Sam and Massa, Francisco and Lerer, Adam and Bradbury, James and Chanan, Gregory and Killeen, Trevor and Lin, Zeming and Gimelshein, Natalia and Antiga, Luca and others},
  journal={Advances in neural information processing systems},
  volume={32},
  year={2019}
}

@inproceedings{kwon2023efficient,
  title={Efficient memory management for large language model serving with pagedattention},
  author={Kwon, Woosuk and Li, Zhuohan and Zhuang, Siyuan and Sheng, Ying and Zheng, Lianmin and Yu, Cody Hao and Gonzalez, Joseph and Zhang, Hao and Stoica, Ion},
  booktitle={Proceedings of the 29th symposium on operating systems principles},
  pages={611--626},
  year={2023}
}

@inproceedings{Triton2019,
  title={Triton: an intermediate language and compiler for tiled neural network computations},
  author={Tillet, Philippe and Kung, Hsiang-Tsung and Cox, David},
  booktitle={Proceedings of the 3rd ACM SIGPLAN International Workshop on Machine Learning and Programming Languages},
  pages={10--19},
  year={2019}
}

@inproceedings{deepspeed2020,
  title={Deepspeed: System optimizations enable training deep learning models with over 100 billion parameters},
  author={Rasley, Jeff and Rajbhandari, Samyam and Ruwase, Olatunji and He, Yuxiong},
  booktitle={Proceedings of the 26th ACM SIGKDD international conference on knowledge discovery \& data mining},
  pages={3505--3506},
  year={2020}
}

@article{zheng2024sglangefficientexecutionstructured,
  title={Sglang: Efficient execution of structured language model programs},
  author={Zheng, Lianmin and Yin, Liangsheng and Xie, Zhiqiang and Sun, Chuyue Livia and Huang, Jeff and Yu, Cody Hao and Cao, Shiyi and Kozyrakis, Christos and Stoica, Ion and Gonzalez, Joseph E and others},
  journal={Advances in neural information processing systems},
  volume={37},
  pages={62557--62583},
  year={2024}
}

@article{li2025autotritonautomatictritonprogramming,
  title={Autotriton: Automatic triton programming with reinforcement learning in llms},
  author={Li, Shangzhan and Wang, Zefan and He, Ye and Li, Yuxuan and Shi, Qi and Li, Jianling and Hu, Yonggang and Che, Wanxiang and Han, Xu and Liu, Zhiyuan and others},
  journal={arXiv preprint arXiv:2507.05687},
  year={2025}
}

@inproceedings{zhu2026qimeng,
  title={Qimeng-kernel: Macro-thinking micro-coding paradigm for llm-based high-performance gpu kernel generation},
  author={Zhu, Xinguo and Peng, Shaohui and Guo, Jiaming and Chen, Yunji and Guo, Qi and Wen, Yuanbo and Qin, Hang and Chen, Ruizhi and Zhou, Qirui and Gao, Ke and others},
  booktitle={Proceedings of the AAAI Conference on Artificial Intelligence},
  volume={40},
  number={34},
  pages={29168--29176},
  year={2026}
}

@misc{kernelfalcon2025,
  author       = {{PyTorch Team and Contributors}},
  title        = {KernelFalcon: Autonomous GPU Kernel Generation via Deep Agents},
  year         = {2025},
  howpublished = {\url{https://github.com/meta-pytorch/kernelagent}},
  note         = {Accessed: 2025-07}
}

@inproceedings{li2025tritonbenchbenchmarkinglargelanguage,
  title={Tritonbench: Benchmarking large language model capabilities for generating triton operators},
  author={Li, Jianling and Li, Shangzhan and Gao, Zhenye and Shi, Qi and Li, Yuxuan and Wang, Zefan and Huang, Jiacheng and WangHaojie, WangHaojie and Wang, Jianrong and Han, Xu and others},
  booktitle={Findings of the Association for Computational Linguistics: ACL 2025},
  pages={23053--23066},
  year={2025}
}

@inproceedings{ouyangkernelbench,
  title={KernelBench: Can LLMs Write Efficient GPU Kernels?},
  author={Ouyang, Anne and Guo, Simon and Arora, Simran and Zhang, Alex L and Hu, William and Re, Christopher and Mirhoseini, Azalia},
  booktitle={Forty-second International Conference on Machine Learning},
  year={2025}
}

@article{xing2026flashinfer,
  title={Flashinfer-bench: Building the virtuous cycle for ai-driven llm systems},
  author={Xing, Shanli and Zhai, Yiyan and Jiang, Alexander and Dong, Yixin and Wu, Yong and Ye, Zihao and Ruan, Charlie F and Huang, Yingyi and Zhang, Yineng and Yin, Liangsheng and others},
  journal={Proceedings of Machine Learning and Systems},
  volume={8},
  pages={2016--2064},
  year={2026}
}

@online{reddit_sakana_cheating_2025,
  author  = {{Reddit Community}},
  title   = {Sakana discovered its AI CUDA engineer cheating},
  year    = {2025},
  url     = {https://www.reddit.com/r/OpenAI/comments/1iwc24f/sakana_discovered_its_ai_cuda_engineer_cheating/},
  urldate = {2025-03-01}
}

@article{liu2026dr,
  title={Dr. Kernel: Reinforcement Learning Done Right for Triton Kernel Generations},
  author={Liu, Wei and Xu, Jiawei and Li, Yingru and Zheng, Longtao and Li, Tianjian and Liu, Qian and He, Junxian},
  journal={arXiv preprint arXiv:2602.05885},
  year={2026}
}

@article{hammond2026agentic,
  title={Agentic operator generation for ml asics},
  author={Hammond, Alec and Markosyan, Aram and Dontula, Aman and Mahns, Simon and Fisches, Zacharias and Pedchenko, Dmitrii and Muzumdar, Keyur and Supper, Natacha and Cao, Site and Zhu, Haishan and others},
  journal={Proceedings of Machine Learning and Systems},
  volume={8},
  pages={1583--1594},
  year={2026}
}

@article{du2025akg,
  title={AKG kernel Agent: A Multi-Agent Framework for Cross-Platform Kernel Synthesis},
  author={Du, Jinye and Yuan, Quan and Zhang, Zuyao and Yi, Yanzhi and Hu, Jiahui and Chen, Wangyi and Zhu, Yiyang and Zheng, Qishui and Zou, Wenxiang and Chang, Xiangyu and others},
  journal={arXiv preprint arXiv:2512.23424},
  year={2025}
}

@misc{fisches2025kernelllm,
  title  = {KernelLLM: Making Kernel Development More Accessible},
  author = {Fisches, Zacharias V. and Paliskara, Sahan and Guo, Simon and Zhang, Alex and Spisak, Joe and Cummins, Chris and Leather, Hugh and Synnaeve, Gabriel and Isaacson, Joe and Markosyan, Aram and Saroufim, Mark},
  url    = {https://huggingface.co/facebook/KernelLLM},
  year   = {2025}
}

@article{woo2025tritonrl,
  title={Tritonrl: Training llms to think and code triton without cheating},
  author={Woo, Jiin and Zhu, Shaowei and Nie, Allen and Jia, Zhen and Wang, Yida and Park, Youngsuk},
  journal={arXiv preprint arXiv:2510.17891},
  year={2025}
}

@article{xia2025live,
  title={Live-SWE-agent: Can Software Engineering Agents Self-Evolve on the Fly?},
  author={Xia, Chunqiu Steven and Wang, Zhe and Yang, Yan and Wei, Yuxiang and Zhang, Lingming},
  journal={arXiv preprint arXiv:2511.13646},
  year={2025}
}

@article{gao2025trae,
  title={Trae agent: An llm-based agent for software engineering with test-time scaling},
  author={Gao, Pengfei and Tian, Zhao and Meng, Xiangxin and Wang, Xinchen and Hu, Ruida and Xiao, Yuanan and Liu, Yizhou and Zhang, Zhao and Chen, Junjie and Gao, Cuiyun and others},
  journal={arXiv preprint arXiv:2507.23370},
  year={2025}
}

@misc{huang2026realistictritonbench,
  author = {Huang, Jinjun and Wen, Zhongzhen and Xu, Tongtong and Yan, Meng and Xia, Xin and Liu, Zhongxin},
  title  = {Online Appendix for ``RealisticTritonBench: A Benchmark for Triton-Kernel Generation in Real-World AI Frameworks''},
  year   = {2026},
  url    = {https://anonymous.4open.science/r/RealisticTritonBench-2583/appendix/appendix.md}
}

@inproceedings{jiang2025issue,
  title={Issue Localization via LLM-Driven Iterative Code Graph Searching},
  author={Jiang, Zhonghao and Ren, Xiaoxue and Yan, Meng and Jiang, Wei and Li, Yong and Liu, Zhongxin},
  booktitle={2025 40th IEEE/ACM International Conference on Automated Software Engineering (ASE)},
  pages={3034--3045},
  year={2025},
  organization={IEEE}
}

@article{wang2026icore,
  title={iCoRe: An Iterative Correlation-Aware Retriever for Bug Reproduction Test Generation},
  author={Wang, Junyi and Cao, Jialun and Liu, Zhongxin},
  journal={Proceedings of the ACM on Software Engineering},
  volume={3},
  number={FSE},
  pages={4231--4252},
  year={2026},
}

\end{document}